\documentclass[%
 reprint,
 amsmath,amssymb,
 aps,
]{revtex4-1}

\usepackage{graphicx}
\usepackage{dcolumn}
\usepackage{bm}

\begin{document}


\title{Molecular dynamics simulations of \\ sodium nanoparticle deposition on magnesium oxide}

\author{Yannick Fortouna$^{1,4}$}
\author{Pablo de Vera$^{1,2}$}%
 \email{pablo.vera@um.es}
 
\author{Alexey Verkhovtsev$^{1,3}$}

\author{Andrey V. Solov'yov$^{1,3}$}
 
\affiliation{%
 $^{1}$MBN Research Center, Altenh\"oferallee 3, 60438 Frankfurt am Main, Germany
}%

\affiliation{%
 $^{2}$Departamento de F\'isica--Centro de Investigaci\'on en \'Optica y Nanof\'isica, Regional Campus of International Excellence “Campus Mare Nostrum”, Universidad de Murcia, 30100 Murcia, Spain
}%

\affiliation{%
 $^{3}$On leave from A. F. Ioffe Physical-Technical Institute, Polytekhnicheskaya 26, 194021 St. Petersburg, Russia
}%

\affiliation{%
 $^{4}$Currently at Department of Materials Science and Engineering,
University of Ioannina, 45110 Ioannina, Greece
}%





\begin{abstract}
The interaction of mass-selected atomic clusters and nanoparticles with surfaces attracts strong interest in view of fundamental research and technological applications. 
Understanding dynamics of the deposition process is important for controlling structure and functioning of deposited nanoparticles on a substrate, but experimental techniques can usually observe only the final outcome of the deposition process.
In this paper, the deposition of 4 nm-sized sodium nanoparticles on an experimentally relevant magnesium oxide substrate is studied by means of classical molecular dynamics simulations. 
An empirical force field is derived which accounts for the interaction of highly polarizable Na atoms with the surface, reproducing the results of previously reported quantum mechanics/molecular mechanics simulations. Molecular dynamics simulations permit exploring the dynamics of deposited nanoparticles on long timescales on the order of hundreds of picoseconds, thus enabling the analysis of energy relaxation mechanisms and the evolution of nanoparticle structure up to its thermalization with the substrate. Several nanoparticle characteristics, such as internal structure, contact angle, and 
aspect ratio
are studied in a broad deposition energy range from the soft landing to multi-fragmentation regimes. 
\end{abstract}

\maketitle


\section{Introduction}

Metal clusters, nanoparticles and nanoalloys have been a subject of intense research over the past decades \cite{Heer1993, Haberland1994, Kreibig_Vollmer, MeiwesBroer2000, ISACC_LatestAdv_2008, Suraud_ClusterScience, Calvo_2015_PCCP.17.27922, ComputModel_Nanomaterials}.
Unique and size-dependent structural, electronic, optical and magnetic properties of these systems have led to various technological applications.
For instance, metal clusters and nanoparticles, both monatomic and bimetallic, can be used as junctions in nanoelectronic devices \cite{Janes2000}
or as elements of photonic crystals \cite{Sun2018}. They are also useful for energy, environmental and medical applications,
e.g. as catalysts \cite{Heiz2008}, contrast agents in medical imaging \cite{NPs_for_imaging}, and radiosensitizers in cancer treatment with ionizing radiation \cite{Haume2016a, Verkhovtsev_2015_PRL.114.063401, bimetallic_radiosensitizers}.

\begin{sloppypar}
Many of these applications involve the interaction of clusters with molecular environments or with surfaces which serve as a support \cite{MeiwesBroer2000}. Understanding the dynamics of cluster deposition is of high relevance for controlling the structure and properties of supported clusters. Depending on deposition conditions, structure of clusters on the substrate can be either preserved or changed substantially, or clusters may experience fragmentation including possible penetration into the substrate and/or modification of the latter. The ability to control these processes lays in the core of key experimental techniques for the fabrication of thin films and nanodevices, such as cluster ion beam assisted deposition, sputtering, surface smoothing or substrate implantation \cite{Popok2011, ClusterBeamDeposition_book_2020}.
\end{sloppypar}

\begin{sloppypar}
The shape of clusters and nanoparticles on a surface is determined by the interplay of different processes and phenomena, including electron shell closure \cite{Poenary_2007_EPL.79.63001, Poenaru_2008_EPJD.47.379, Haekkinen_2016_AdvPhysX.1.467}, interaction of deposited systems with the substrate \cite{Hoevel_1998_PRL.81.4608, Verkhovtsev2020_clusters}, relaxation of thermal energy remaining after the collision by means of heat transfer, and atomic rearrangements caused by collision-induced mechanical stress of phase transitions \cite{ISACC_LatestAdv_2008, Suraud_ClusterScience}.
%
%
These processes depend on the type and temperature of the substrate, the size and composition of the deposited cluster/nanoparticle and on the deposition energy. 
For small atomic clusters containing $N \lesssim 200$ atoms quantum effects, such as even-odd oscillations in cluster abundance spectra and the appearance of ``magic'' numbers associated with electron shell closure, play a crucial role in determining the shape of clusters on a surface \cite{Poenaru_2008_EPJD.47.379}. However, these effects shrink with increasing the system size up to $N \sim 10^3$ \cite{Suraud_ClusterScience}.
While clusters made up of a few atoms usually keep their structure upon soft landing (i.e. when kinetic energy per atom is much smaller than the cluster cohesion energy) \cite{Hakkinen1996,Bar2008}, large clusters and nanoparticles can experience significant deformations, such as flattening, surface wetting or epitaxial alignment \cite{Popok2011}. Hard deposition at kinetic energies exceeding the cohesion energy can lead to cluster fragmentation.
\end{sloppypar}

\begin{sloppypar}
The deposition of metal clusters and nanoparticles on various surfaces has been widely studied experimentally (see e.g. Ref.~\cite{Popok2011} and references therein). However, experimental studies are usually limited to the observation of the final state of the deposition process, and the study of the cluster deposition mechanisms thus commonly relies on theoretical methods \cite{Dinh2010,Popok2011}. For instance, detailed quantum mechanics/molecular mechanics (QM/MM) simulations of the deposition of small sodium clusters on magnesium oxide substrates were reported in Refs.~\cite{Bar2007,Bar2008,Bar2009,Dinh2010}. These materials were selected to highlight the importance of both electron and nuclear dynamics in the process of cluster deposition. Sodium is a highly polarizable metal having a single valence electron. Magnesium oxide is a hard ionic crystal with a highly corrugated potential energy surface which can easily polarize Na atoms and produce complex interaction patterns.

The study of metallic aggregates deposited onto oxide surfaces is very important for technological applications, especially in the field of catalysis \cite{Henry_1998_SurfSciRep.31.231}. Deposition, dynamics and diffusion of metal clusters on MgO films has particularly attracted both experimental and theoretical interest \cite{Henry_1998_SurfSciRep.31.231, Barcaro_2005_PRL.95.246103, Barcaro_2005_JCTC.1.972}, including a number of recent studies \cite{Paradiso_2020_JPCC.124.14564, Buendia_2020_JCP.152.024303, Magkoev_2020_RusJPCA.94.401, Nigam_2020_ApplSurfSci.506.144963}. 

QM/MM simulations conducted in Refs.~\cite{Bar2007,Bar2008,Bar2009,Dinh2010} explored different deposition regimes for small Na$_6$ and Na$_8$ clusters including soft landing, hard collision and reflection of the clusters. 
Structure and dynamics of the clusters as well as the mechanisms of energy transfer to the substrate were analyzed in detail. However, due to the complexity and computational cost of QM/MM calculations, this analysis could only be carried out for very small clusters, short simulation times (on the order of several picoseconds) and zero temperature. 
\end{sloppypar}

Contrary to \textit{ab initio} approaches, classical molecular dynamics (MD) permits simulating deposition of much bigger systems at finite temperature and over significantly longer timescales, thus allowing to explicitly account for the process of energy relaxation \cite{Averback1994,Carroll2000,Colla2000,Hou2000,Moseler2000, Verkhovtsev2020_clusters}. Empirical force fields employed in MD simulations can be tuned to effectively reproduce 
the collision dynamics observed in QM/MM calculations on much shorter time scales.

\begin{sloppypar}
This paper reports a detailed theoretical analysis of the deposition of Na$_{1067}$ nanoparticles ($\sim 4$ nm in diameter) on a MgO (001) substrate. The simulations were conducted for a wide deposition energy range (from the soft-landing to nanoparticle multifragmentation regimes) and at several finite temperatures at which the nanoparticle is either in the solid state or in the form of a liquid droplet. 
A force field describing the interaction of Na atoms with the MgO surface was developed and used in the simulations.
The force field reproduced main features of the deposition of a single Na atom, obtained previously in QM/MM simulations \cite{Bar2008}. The MD technique was then employed to simulate deposition and post-collision dynamics of the nanometer-sized sodium nanoparticles.
Several aspects of the deposition process are explored as a function of deposition energy, including the change of the nanoparticle structure and shape, its wetting properties, as well as the energy transfer between the nanoparticle and the substrate. 
This study provides detailed atomistic-level insights into the dynamics of sodium nanoparticle deposition on magnesium oxide substrates, which complement the information already gathered for small clusters from QM/MM simulations and which may be useful for experimental studies.
\end{sloppypar}

\section{Computational methodology}
\label{sec:methods}

In the performed classical MD simulations the coupled Langevin equations for all atoms in the system were solved numerically by means of the leapfrog algorithm \cite{Solovyov2017b}. 
Simulations of the deposition of a 4 nm diameter sodium nanoparticle on MgO(001) surface were performed for systems pre-equilibrated at 77~K and 300~K. 
These temperatures allow the comparison of the deposition dynamics for a solid nanoparticle and a liquid droplet, having into account that the melting temperature of Na$_{1067}$ is slightly below 300~K \cite{Martin1994}.

\begin{sloppypar}
All simulations were performed by means of MBN Explorer \cite{Solovyov2012}, a software package for the advanced multiscale modeling of complex molecular structure and dynamics. The MgO substrate and the sodium nanoparticle were constructed by means of its dedicated graphical user interface, MBN Studio \cite{Sushko2019}. This software was used also to prepare all other necessary input files and to analyze simulation results. In the following sections, the construction and preparation of each part of the system is described, together with the potential used for each interatomic interaction.
\end{sloppypar}

\subsection{Magnesium oxide substrate}
\label{sec:MgO}

The interactions involving Mg and O atoms were described based on the nuclear contribution to the empirical potential which was defined in earlier QM/MM calculations \cite{Bar2008,Dinh2010}. Details of this potential,
being a combination of exponential and power potentials,
as well as its parameters can be found in Table~4 of Ref.~\cite{Dinh2010}.
All atoms in the substrate carried partial charges of $\pm 2|e|$ (with $e$ being the elementary charge) and thus interacted also through the Coulomb potential. The electrostatic interactions were treated by means of the Ewald algorithm implemented in MBN Explorer \cite{Solovyov2017b}.

\begin{sloppypar}
A face-centered cubic structure of MgO with a lattice parameter of 4.212~\AA~was employed to create substrates with the size of $12.21 \times 12.21$ nm$^2$ and $24.43 \times 24.43$ nm$^2$ in the $x$-$y$ directions, simulated using periodic boundary conditions.  
Following Ref.~\cite{Dongare2005} the substrate was formed by seven atomic layers in the $z$-direction normal to the surface. The height of the constructed substrate (12.6 \AA) exceeded significantly the range of Na--Mg and Na--O interatomic interactions (see next subsection).
\end{sloppypar}

The structure of MgO was optimized using the velocity quenching algorithm and the time step of 0.1~fs. After structure optimization the substrate was equilibrated using the Langevin thermostat with a damping time of 0.1~ps to the target temperatures of 77 K and 300 K 
such that obtained atomic velocities corresponded to the Maxwell-Boltzmann distribution.







\subsection{Sodium atom--substrate interaction}
\label{sec:NaMgO}

\begin{sloppypar}
The interaction between Na and O or Mg atoms was described by means of the following pairwise potential:
\begin{eqnarray}
U_{ij}(r) &=& D_{ij}  \left[  e^{{\displaystyle -2 \beta_{ij} \left( r - r_{0,ij} \right)} }-
                               2 e^ {{\displaystyle -\beta_{ij} \left( r - r_{0,ij} \right)} } \right]  \nonumber \\
                                &-&   {\displaystyle\frac{C_{ij}}{r^{4}}  + \frac{q_i q_j}{{\displaystyle\varepsilon_0  r}}} \, \mbox{,} 
				\label{MOrse_Polarization}
\end{eqnarray}
where $r$ is the distance between atoms $i$ and $j$. The first term on the r.h.s. describes a Morse-type interaction, whereas the second and third terms describe the long-range polarization and electrostatic interactions, respectively. Parameters $D_{ij}$, $r_{0,ij}$ and $\beta_{ij}$ of the Morse potential represent the depth of the potential well,  equilibrium interatomic distance and steepness of the potential for each pair of atoms. $q_i$ and $q_j$ are atomic partial charges, $\varepsilon_0$ is the effective charge screening factor (set equal to 1 in present simulations) and $C_{i}$ is an empirical parameter determining the intensity of the polarization forces for the interaction of Na with a particular atom $i$. The long-range Coulomb interaction was calculated by means of the Ewald algorithm \cite{Solovyov2017b}. The partial charge $q_{\rm Na}$ on the Na atom was treated as an additional free parameter to account for additional attracting forces between the atom and the surface due to its strong polarizability in the field of the ionic crystal.
\end{sloppypar}



\begin{figure}[t]
\centering
\includegraphics[width=\columnwidth]{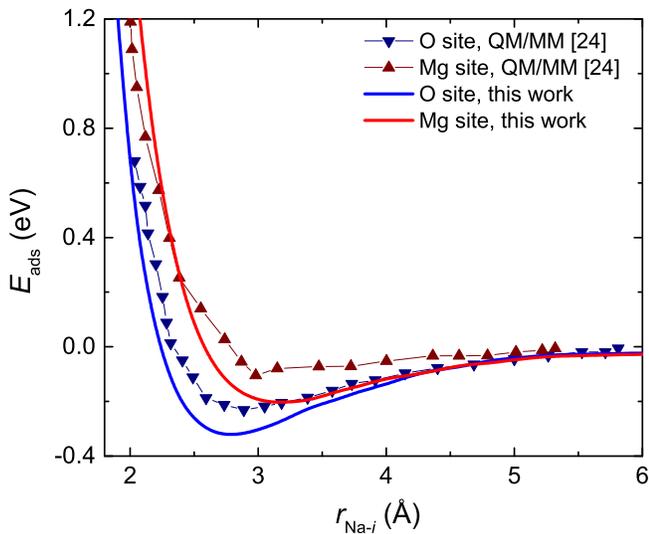}
\caption{
Adsorption energy of a Na atom on the MgO surface as a function of the distance between Na and O or Mg sites. 
Present MD 
empirical force field calculations (lines) are compared to QM/MM results (symbols) \cite{Bar2008}.
}
\label{fig_FF}
\end{figure}

The parameters for the potential (\ref{MOrse_Polarization}) were obtained by a trial and error procedure in an iterative manner until a reasonable agreement with reference data from QM/MM simulations \cite{Bar2008,Dinh2010} has been reached. In particular, the adsorption energy and dynamics of a single Na atom on top of Mg and O sites on the surface of MgO were evaluated and compared with the reference \textit{ab initio} calculations \cite{Bar2008,Dinh2010}. 
The best agreement in terms of both energetics and dynamics of a Na atom atop MgO 
was achieved with the parameters summarized in Table~\ref{table:FF_parameters}. The resulting adsorption energy curves for a Na atom on top of Mg and O sites are shown in Fig.~\ref{fig_FF}. A partial charge $q_{\rm Na} = +0.29|e|$ was assigned to the Na atom in this case.
Note that, when the Na--O and Na--Mg interactions were described only with the Morse potential or when the partial charge on a sodium atom was set equal to zero, simulated MD trajectories deviated from the earlier QM/MM results, which can be attributed to underestimation of the attractive forces.
The calculated adsorption energies reproduce the shape of the QM/MM results, presenting only slightly deeper potential wells. It should be noted that variation on the order of a few tenths of eV can arise by considering different exchange-correlation functionals and basis sets in quantum-chemistry calculations. 
We have ensured that the derived empirical potential, Eq.~(\ref{MOrse_Polarization}), provides a reasonable agreement with the dynamics of a single Na atom on MgO and gives results consistent with QM/MM simulations of Na$_6$ and Na$_8$ cluster deposition \cite{Bar2008}. The constructed potential is thus deemed suitable for the simulation of the deposition of a nanometer-sized sodium nanoparticle.


\begin{table}[t]
	\centering
		\caption{Parameters of the force field for Na--O and Na--Mg interactions, Eq.~(\ref{MOrse_Polarization}), used in the simulations.}
\begin{tabular}{cc c cc }
\hline

   Atom pair               &  $D_{ij}$~(eV)  & $\beta_{ij}$ ({\AA$^{-1}$})  &  $r_{0,ij}$~(\AA) & $C_i$~(eV)    \\
\hline

Na--O            &  0.099  & 1.5 &  2.94 &  0.5  \\

Na--Mg           &  0.001  & 1.32 &  4.85 &  0.5  \\

	\end{tabular}
		\label{table:FF_parameters}
		\end{table}




\subsection{Sodium nanoparticle}
\label{sec:Nan}

A spherical sodium nanoparticle of 4 nm diameter, containing 1067 atoms, was cut out from the corresponding bulk crystal by means of MBN Studio \cite{Sushko2019}. Geometry of the nanoparticle was first optimized using the velocity quenching algorithm with a 0.1~fs time step. The many-body Gupta potential was used to describe the interatomic interactions with parameters taken from Ref.~\cite{Li1997}. 

\begin{figure}[t]
\centering
\includegraphics[width=0.5\textwidth]{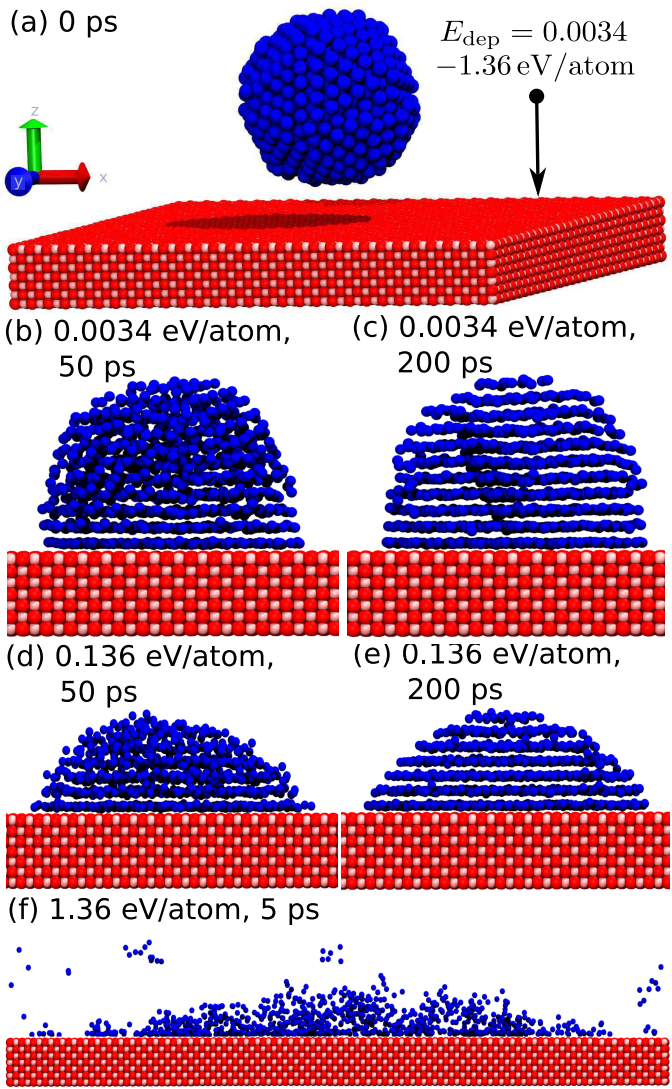}
\caption{
MD snapshots of the Na$_{1067}$ nanoparticle deposited on a MgO substrate at several deposition energies. Panel (a) shows the initial geometry of the system. Panels (b-f) illustrate three different deposition regimes, see text for details: (b)-(c) nanoparticle structure after 50 and 200~ps, respectively, for $E_{\rm dep} =  0.0034$~eV/atom; (d)-(e) the same as previous for $E_{\rm dep} = 0.136$~eV/atom; (f) snapshot of the fragmented nanoparticle deposited at $E_{\rm dep} = 1.36$~eV/atom after 5~ps.}
\label{fig_trajectories}
\end{figure}

\begin{sloppypar}
After initial energy minimization the nanoparticle was annealed to create a more energetically favorable starting geometry for the deposition simulations. Several annealing cycles were simulated following the procedures reported in Ref.~\cite{Ellaby2018}. The first cycle consisted of heating from 0~K to 400~K at a rate of 0.08~K/ps, followed by a constant temperature simulation at 400~K for 2~ns, and cooling down to 0~K at a rate of 0.08 K/ps. The follow-up cycles were similar but the nanoparticle was heated up to 200~K, a temperature slightly below the cluster melting temperature, to allow surface reorganization without complete melting of the nanoparticle. 
The nanoparticle melting temperature of 260~K was determined by simulating heating of the annealed structure. The evaluated melting temperature of the nanoparticle is close to the experimentally determined values of $280-290$~K \cite{Martin1994}.
Cohesive energy of the nanoparticle converged to 1.0194~eV/atom after three annealing cycles. 
A nearly icosahedral shape was obtained (see Fig.~\ref{fig_trajectories}(a)), comprising a mix of face centered cubic and hexagonal compact structures, in accordance with the structures known for sodium clusters \cite{Haberland1994}. The nanoparticle was equilibrated to the target temperatures of 77~K and 300~K by means of the Langevin thermostat and the resulting velocities corresponded to the Maxwell-Boltzmann distribution.
\end{sloppypar}

\subsection{Deposition simulations}
\label{sec:deposition}

MD simulations of the nanoparticle deposition on the substrate were performed in the $NVE$ microcanonical ensemble, thus ensuring conservation of the total energy of the system. The nanoparticle was placed in the center of the simulation box at a 10 \AA~distance from the surface, such that initial nanoparticle--substrate interactions were negligible (see Fig.~\ref{fig_FF}). Initial atomic velocities were taken from pre-equilibration simulations at 77~K and 300~K in order to simulate the deposition process at low temperature at which the nanoparticle is solid, as well as at room temperature at which the nanoparticle has the shape of a liquid droplet. The partial charge of $+0.29|e|$, derived in the fitting procedure described in Sect.~\ref{sec:NaMgO}, was equally distributed among all the atoms in the nanoparticle.
Additional velocities in the direction normal to the surface were given to every atom of the nanoparticle such that the nanoparticle was deposited with kinetic energies $E_{\rm dep} = 
0.0034$, 
0.0068, 
0.0136,
0.034, 
0.068, 
0.102, 
0.136, 
0.34, 
0.68 
and 
1.36 eV/atom.
%
The initial configuration of the system is illustrated in Fig.~\ref{fig_trajectories}(a).

A $12.21 \times 12.21$~nm$^2$ substrate was used for most of the simulations. However, a larger substrate of $24.43 \times 24.43$~nm$^2$ was employed for deposition energies larger than 0.34~eV/atom to avoid interaction of the heavily deformed or fragmented nanoparticle with its periodic images. Two bottom MgO layers were fixed to avoid the displacement of the substrate upon nanoparticle impact. Simulations were performed using the leapfrog algorithm 
with a 1~fs time step, 
which ensured that variation of the total energy did not exceed 0.1\%. As described in Section~\ref{sec:results}, most of the phenomena arising during deposition have been observed within the first 50~ps of the simulations. Nonetheless, longer simulations up to 500~ps were conducted in some cases to analyze the dynamics of the system on the longer time scale.

\section{Results and discussion}
\label{sec:results}

In subsection \ref{sec:shape} we briefly discuss the time evolution of the shape and structure of the nanoparticle at different deposition energies. 
Then, in subsection \ref{sec:therm} we analyze how fast the nanoparticle has reached thermal equilibrium with the substrate after the deposition. In subsection \ref{sec:results_contact_angle} we quantitatively characterize the final shape acquired by the nanoparticle at different deposition energies and evaluate the contact angle with the substrate. Finally, the longer-term changes of the internal structure of the nanoparticle are analyzed in subsection \ref{sec:results_RDF}.

\subsection{Shape and structure of the deposited nanoparticle}
\label{sec:shape}

\begin{sloppypar}
Figure~\ref{fig_trajectories} presents several MD snapshots of the Na$_{1067}$ nanoparticle (pre-equilibrated at 77~K) deposited on MgO at different deposition energies $E_{\textrm{dep}}$. Three distinct deposition regimes have been observed as shown in Fig.~\ref{fig_trajectories}(b-f). 
In the ``soft'' collision regime (e.g. at $E_{\rm dep} = 0.0034$~eV/atom, see panels (b) and (c)) the nanoparticle remains in the solid phase in the course of deposition. At more energetic collisions (e.g. at $E_{\rm dep} = 0.136$~eV/atom, see panels (d) and (e)) the nanoparticle undergoes a collision-induced melting phase transition followed by its subsequent re-crystallization. Finally, the ``hard'' collision regime (e.g. at $E_{\rm dep} = 1.36$~eV/atom, panel (f)) leads to rapid multifragmentation of the nanoparticle. A detailed analysis of the nanoparticle shape is presented below in Section~\ref{sec:results_contact_angle}.
\end{sloppypar}

\begin{sloppypar}
Figures~\ref{fig_trajectories}(b-e) show snapshots from the simulations taken at time instances of 50~ps and 200~ps. The shape of the nanoparticle is stabilized by about 50~ps and does not change significantly at larger simulation times. The nanoparticle deposited at $E_{\rm dep} = 0.0034$~eV/atom acquires the shape of a truncated prolate spheroid, see panels (b) and (c). Deposition at the 40 times larger energy results in the formation of a truncated oblate spheroid as shown in panels (d) and (e). 
Note that some sodium atoms arrange into layers parallel to the MgO surface as a result of the nanoparticle--substrate interaction; the formation of two sodium layers is clearly seen after the first 50~ps of the simulations while the whole nanoparticle acquires a layered structure after 200~ps. 
This behavior is similar to the well-known phenomenon of epitaxial alignment which was studied both experimentally and computationally \cite{Popok2011}.
Finally, Fig.~\ref{fig_trajectories}(f) illustrates the case of ``hard'' deposition leading to rapid fragmentation of the nanoparticle. At deposition energy of 1.36~eV/atom (which exceeds the calculated cohesive energy of the nanoparticle, 1.019~eV/atom) multi-fragmentation of Na$_{1067}$ and the formation of single sodium atoms and small clusters was observed within the first 5~ps of the simulation. 
\end{sloppypar}

\subsection{Thermalization of the nanoparticle on the surface}
\label{sec:therm}

\begin{figure}[t]
\centering
\includegraphics[width=0.85\columnwidth]{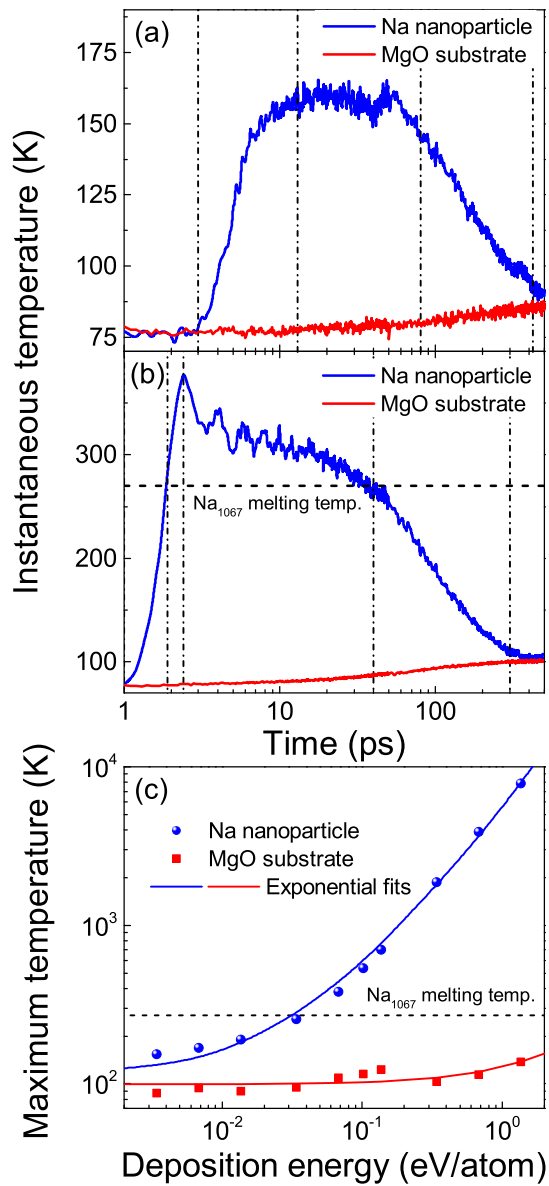}
\caption{
Instantaneous temperature of the Na$_{1067}$ nanoparticle and the MgO substrate as a function of simulation time for deposition energies of (a) 0.0068~eV/atom and (b) 0.068 eV/atom. Panel (c) shows maximum instantaneous temperature of the nanoparticle and the substrate as a function of deposition energy.
}
	\label{fig_Temp}
\end{figure}

\begin{sloppypar}
Figure~\ref{fig_Temp} illustrates how instantaneous temperatures of the nanoparticle and the substrate evolved in the course of simulations. Results shown in panels (a) and (b) were obtained for the nanoparticle pre-equilibrated at 77~K and deposited at $E_{\textrm{dep}} = 0.0068$~eV/atom and 0.068~eV/atom, respectively. As shown in Figure~\ref{fig_Temp}, the nanoparticle reached thermal equilibrium with the substrate at time instances of about 400--500~ps. Before thermalization, temperature of the nanoparticle exhibited a similar behavior in the two cases considered: it first increased rapidly as the nanoparticle hit the surface and then it decreased gradually while the nanoparticle transferred its energy to the substrate. 
Figure~\ref{fig_Temp}(c) shows the maximum temperatures of the nanoparticle and the substrate, reached after the first 50~ps of the simulation, for different deposition energies considered in this study. The calculated dependencies of the maximal temperatures $T_{\textrm{max}}$ on $E_{\textrm{dep}}$ were fitted with an exponential function:
\end{sloppypar}
\begin{equation}
T_{\rm max}  =  T_0 + T_1 \, e^{\omega \, E_{\rm dep} } \ , 
\label{eq:fitT}
\end{equation}
where $T_0$, $T_1$ and $\omega$ are the fitting parameters listed in Table~\ref{table:Fit_parameters_Tmax_AR_WCA}.

\begin{table}[t]
	\centering
		\caption{Parameters for the fitting functions, Eqs.~(\ref{eq:fitT}), (\ref{eq:exp_fit_delta}) and (\ref{eq:exp_fit_theta}), describing, respectively, the maximum instantaneous temperatures of the nanoparticle and the substrate after deposition,
	  the nanoparticle aspect ratio
		and the contact angle between the nanoparticle and the substrate at the end of simulations.}
\begin{tabular}{ccc}
\hline
                       &  Na$_{1067}$  &  MgO         \\
\hline
$T_{0}$~(K)            &  $-16640.6$     &  $-40743.0$   \\
$T_{1}$~(K)            &  16754.7      &   40842.4       \\
$\omega$~(eV$^{-1}$/atom)     &  0.28          &  $6.8 \times 10^{-4}$       \\
\hline \hline
$\delta_0$             &  3.52         &    \\
$\delta_1$             &  $-2.18$        &    \\
$\alpha$~(eV$^{-1}$/atom)               &  16.62        &    \\
\hline \hline
$\theta_{0}$~(deg.)    &  68.59        &    \\
$\theta_{1}$~(deg.)    &  23.61        &    \\
$\gamma$~(eV$^{-1}$/atom)     &  33.92        &    \\
\end{tabular}
\label{table:Fit_parameters_Tmax_AR_WCA}
\end{table}

\begin{sloppypar}
Dashed horizontal lines in Fig.~\ref{fig_Temp}(b-c) illustrate the melting temperature of Na$_{1067}$ obtained from simulations, $T_{\textrm{m}} \approx 260$~K. Temperature of the nanoparticle deposited at $E_{\textrm{dep}} = 0.0068$~eV/atom did not exceed that threshold value during the whole simulation (see Fig.~\ref{fig_Temp}(a)), indicating that the nanoparticle remained in the solid phase during deposition. In contrast, instantaneous temperature of the nanoparticle deposited at 0.068~eV/atom exceeds the melting temperature after the first few picoseconds, see Fig.~\ref{fig_Temp}(b). The temperature reaches the maximal value of about 380~K and remains larger than the melting temperature of the nanoparticle for several tens of picoseconds, until it drops below the threshold value, gradually approaching thermal equilibrium with the substrate at time instances of about 400~ps.
\end{sloppypar}

\begin{sloppypar}
Two regimes which do not lead to nanoparticle fragmentation have thus been identified in the simulations. 
Figure~\ref{fig_Temp}(c) shows that at deposition energies below 0.034~eV/atom the nanoparticle remained in the solid phase throughout the whole simulation. At larger values of $E_{\textrm{dep}}$ the instantaneous temperature of the nanoparticle exceeded the melting temperature so that the nanoparticle experienced the melting phase transition followed by re-crystallization. 
Deposition at energies above 0.68~eV/atom led to fission or multifragmentation of the nanoparticle.
\end{sloppypar} 
 

\subsection{Contact angle and 
aspect ratio
of the nanoparticle}
\label{sec:results_contact_angle}

\begin{sloppypar}
Coordinates of sodium atoms, extracted from each simulated MD trajectory, were used to parameterize the shape of the deposited nanoparticle.
\end{sloppypar}

As follows from the simulated trajectories, the sodium nanoparticle is, to a good approximation, radially symmetric with respect to the main axis $z$. In order to characterize the shape of the nanoparticle, cylindrical coordinates ($\rho, z$) were introduced, where:
\begin{equation}
\rho = \sqrt{ \left(x - x_{\rm CM} \right)^2 + \left(y - y_{\rm CM} \right)^2 } \ ,
\end{equation}
with $x_{\rm CM}$ and $y_{\rm CM}$ being $x$- and $y$-projections of the center of mass of the nanoparticle. The $\rho$-axis lies in the MgO surface plane, whereas $z$-axis is perpendicular to the surface.

\begin{figure}[t]
\centering
\includegraphics[width=\columnwidth]{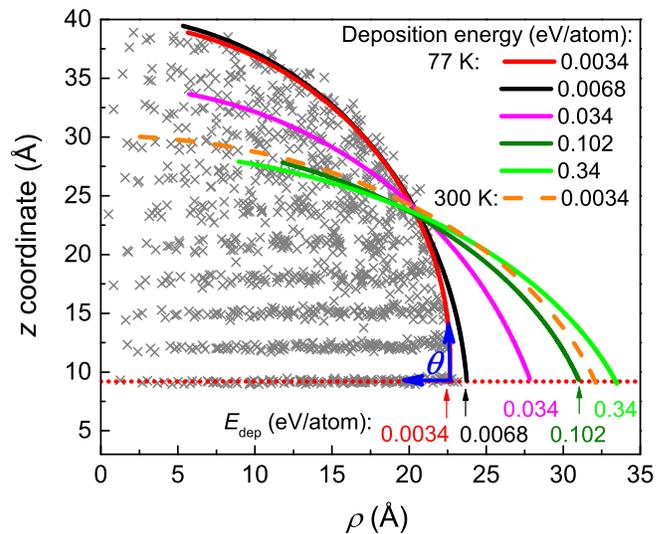}
\caption{
Radial profile of the Na$_{1067}$ nanoparticle deposited on MgO. 
Symbols show the distribution of sodium atoms of the nanoparticle pre-equilibrated at 77~K and deposited at $E_{\textrm{dep}} = 0.0034$~eV/atom, after 50~ps of the simulation. 
Colored curves depict radial profiles of the nanoparticle deposited at different $E_{\textrm{dep}}$, obtained by means of Eq.~(\ref{eq:surface_equation}).
}
\label{fig_NPshapeWCA}
\end{figure}

Figure~\ref{fig_NPshapeWCA} shows by symbols ($z,\rho$) projections of atoms in the Na$_{1067}$ nanoparticle deposited at 77~K with an energy of 0.0034~eV/atom.
The shown distribution of atoms was evaluated at the 50~ps time instance by which the shape of the deposited nanoparticle has been stabilized. The horizontal dotted line of ordinate $z_0 \approx 9$~\AA~depicts average positions of sodium atoms in the bottom-most atomic layer of the nanoparticle, which is closest to the MgO surface.

In order to evaluate the contact angle between the nanoparticle and the substrate, we selected coordinates of atoms located on the nanoparticle's surface and fitted the resulting set of coordinates with the following equation \cite{Verkhovtsev2020_clusters, Giovambattista_2007_JPCB.111.9581, Skvara_2018_MolSimul.44.190}:
\begin{equation}
\rho(z) = \sqrt{ a \left(z - z_0 \right)^2 + b \left(z - z_0 \right) + c}   \quad  , \ \ z \ge z_0 \ , 
\label{eq:surface_equation}
\end{equation}
where $a$, $b$ and $c$ are fitting parameters.
Solid colored curves in Fig.~\ref{fig_NPshapeWCA} show the inverse dependence $z(\rho)$ for the Na$_{1067}$ nanoparticle equilibrated at 77~K and deposited at different energies as indicated by labels. From this dependence the nanoparticle contact angle $\theta$ was determined as \cite{Giovambattista_2007_JPCB.111.9581}: 
\begin{equation}
\theta
= \arctan \left( \left. \frac{dz}{d\rho} \right|_{\rho = \rho^{\prime}} \right) \ ,
\label{eq:contact_angle}
\end{equation}
where $\rho^{\prime}$ is the point of intersection between the fitting curve and the line $z = z_0$ representing the bottom-most Na atomic layer parallel to the substrate surface. 

\begin{sloppypar}
Solid curves shown in Fig.~\ref{fig_NPshapeWCA} illustrate a gradual change of the nanoparticle shape from a truncated prolate spheroid ($z_{\textrm{max}} > \rho_{\textrm{max}}$) to a semi-spheroid ($z_{\textrm{max}} = \rho_{\textrm{max}}$) to a truncated oblate spheroid ($z_{\textrm{max}} < \rho_{\textrm{max}}$) as the deposition energy increases. 
The dashed orange curve in Fig.~\ref{fig_NPshapeWCA} shows the $z(\rho)$ dependence for the nanoparticle equilibrated at 300~K and deposited at $E_{\textrm{dep}} = 0.0034$~eV/atom. It is apparent that the nanoparticle equilibrated at room temperature acquires a droplet-like oblate shape on the surface even at the lowest energy considered.
These results agree with the simulation snapshots shown in Fig.~\ref{fig_trajectories}(b-e).
\end{sloppypar}

Note that the shape of the Na$_{1067}$ nanoparticle equilibrated at 77~K and deposited at 0.34~eV/atom (light green curve) is very similar to that of the nanoparticle equilibrated at 300~K and deposited at much lower energy of 0.0034~eV/atom (dashed orange curve). In the former case the nanoparticle experienced a collision-induced melting phase transition while in the latter case it was directly deposited as a liquid droplet. 
The equilibrium shape of the nanoparticle deposited at 300~K depends very little on deposition energy. As a result, the shapes of the nanoparticles deposited at higher values of $E_{\textrm{dep}}$ (below the threshold energy for nanoparticle fragmentation) are similar to the one shown for $E_{\textrm{dep}} = 0.0034$~eV/atom (see the dashed orange curve). Dependence of the contact angle on $E_{\textrm{dep}}$ is discussed in greater detail below.

\begin{figure}[t]
\centering
\includegraphics[width=0.85\columnwidth]{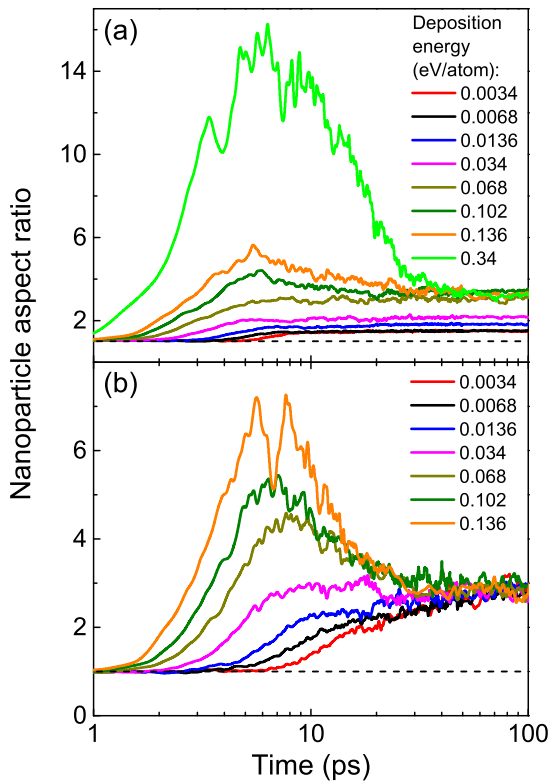}
\caption{
Nanoparticle aspect ratio
$\delta$, Eq. (\ref{eq:aspect_ratio}), as a function of simulation time for deposition at (a) 77~K and (b) 300~K.}
\label{fig_RMSD_AR_t}
\end{figure}

We have also analyzed the ratio of the diameter of the bottom-most atomic layer of the nanoparticle (see red dotted line in Fig.~\ref{fig_NPshapeWCA}) to the nanoparticle height, i.e. the nanoparticle aspect ratio \cite{Popok2011}. It is defined as:
\begin{equation}
\delta = \frac{  2\rho(z_0)  }{  z(\rho = 0)  } \ .
\label{eq:aspect_ratio}
\end{equation}
\begin{sloppypar}
The temporal evolution of $\delta$ for the nanoparticle deposited at different energies is illustrated in Fig.~\ref{fig_RMSD_AR_t}.
Panels~(a) and (b) illustrate the deposition at 77~K and 300~K, respectively.
We consider here the values of $E_{\textrm{dep}}$ up to 0.34~eV/atom for 77~K and up to 0.136~eV/atom for 300K, energies at which the nanoparticle have not fragmented upon collision with the substrate.
Values of $\delta$ between 1 and 2 
indicate that the nanoparticle acquires the shape of a truncated prolate spheroid, whereas $\delta = 2$ describes a perfect semi-spheroid. A parallel can be drawn with the snapshots of nanoparticle deposition at 77~K, shown in Fig.~\ref{fig_trajectories}(b-e). As discussed above, the overall shape of the nanoparticle practically does not change after 50~ps of simulation. This observation is confirmed by constant values of $\delta$ observed after 50~ps. As shown in Fig.~\ref{fig_RMSD_AR_t} this trend has been observed for all deposition energies within the range considered.
\end{sloppypar}

Two distinct deformation regimes have been revealed in the simulations. At high deposition energies ($E_{\textrm{dep}} \ge 0.068$~eV/atom), 
the nanoparticle aspect ratio
increases over the first few picoseconds and eventually converges to the value of $\delta \approx 3$ at the end of simulations for every $E_{\textrm{dep}}$. In contrast, a gradual increase of the final value of $\delta$ from about 1 to 2 is typical for low-energy depositions at energy $E_{\textrm{dep}} \le 0.034$~eV/atom. 

As discussed above, the nanoparticle deposited in the ``soft'' regime remains in the solid phase in the course of deposition; this regime corresponds to a small increase of $\delta$ over time.
When temperature of the nanoparticle exceeds its melting temperature, the nanoparticle experiences the melting phase transition and thus it becomes more susceptible to stress-induced deformation. This occurs at more energetic collisions when the nanoparticle wets the substrate and acquires the shape of an oblate spheroid, see Fig.~\ref{fig_trajectories}(d-e). 
On this basis it is straightforward to explain the dependencies shown in Fig.~\ref{fig_RMSD_AR_t}(b), which describe 
the aspect ratio
for the nanoparticle pre-equilibrated at 300~K. A liquid sodium drop deposited on MgO experiences strong deformation upon contacting the substrate; this behavior is characterized by a rapid increase of $\delta$ (i.e. flattening of the nanoparticle) within the first $5-10$~ps of the simulations. At larger time instances $\delta$ saturates at a constant value of about $3$ when the nanoparticle reaches thermal equilibrium with the substrate. Note that the nanoparticles deposited at different values of $E_{\textrm{dep}}$ have a similar shape as 
the aspect ratio
asymptotically approaches a constant value $\delta \approx 3$.

\begin{figure}[t]
\centering
\includegraphics[width=0.85\columnwidth]{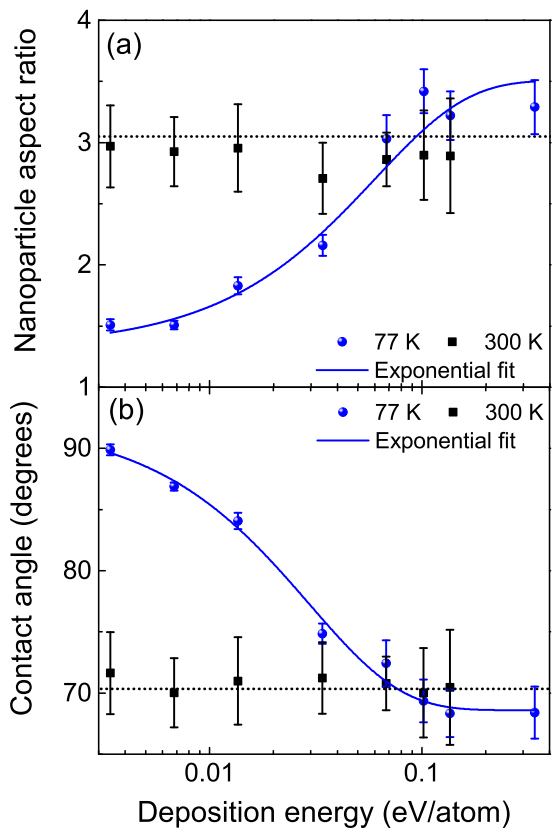}
\caption{
The final aspect ratio (a) and contact angle (b) of the nanoparticle as functions of the deposition energy. Symbols denote time averages, while error bars represent standard deviations. Data for the deposition at 77~K were fitted with exponential functions, Eqs. (\ref{eq:exp_fit_delta}) and (\ref{eq:exp_fit_theta}). 
}
\label{fig_final_charact}
\end{figure}

To complete this analysis, Fig.~\ref{fig_final_charact} presents the comparison of the final 
aspect ratio
and contact angle of deposited nanoparticles, which were initially equilibrated at two different temperatures, 77~K (blue circles) and 300~K (black squares). These parameters were evaluated in a broad deposition energy range below the nanoparticle fragmentation threshold. For each $E_{\textrm{dep}}$, average values and standard deviations in time were calculated for the parts of the trajectories at which $\delta$ and $\theta$ oscillate around a constant value. Symbols represent the mean values while error bars correspond to a standard deviation. 

Two different trends are clearly seen for different deposition temperatures. 
At 300~K (black squares) both 
the aspect ratio
and the contact angle are practically constant within the studied energy range.
Both parameters fluctuate around some characteristic values, $\delta \approx 3$ and $\theta \approx 70 ^{\circ}$, which are indicated by horizontal dotted lines. 
For deposition at 77~K (blue circles), 
aspect ratio
grows exponentially with $E_{\textrm{dep}}$ while the contact angle decreases. 
The dependencies of $\delta$ and $\theta$ on $E_{\textrm{dep}}$ were fitted with the following functions:
\begin{eqnarray}
\label{eq:exp_fit_delta} \delta  &=& \delta_0 + \delta_1 \, e^{\alpha \, E_{\rm dep} } \ , \\
\label{eq:exp_fit_theta} \theta &=& \theta_0 + \theta_1 \, e^{-\gamma \, E_{\rm dep} } \ .
\end{eqnarray}
Parameters of this fit are listed in Table~\ref{table:Fit_parameters_Tmax_AR_WCA}. At deposition energies larger than 0.068~eV/atom both 
aspect ratio
and the contact angle for the nanoparticle deposited at 77~K approach the values obtained at 300~K as a consequence of the collision-induced melting phase transition above this energy. 
These results agree with the conclusions made above that the resulting shape of a liquid droplet deposited at 300~K depends very little on the deposition energy. In contrast, structure and contact angle for the nanoparticle deposited at the lower temperature strongly depend on $E_{\textrm{dep}}$.




\subsection{Collision-induced structural and phase transformations}
\label{sec:results_RDF}

\begin{figure}[t]
\centering
\includegraphics[width=0.85\columnwidth,scale=2]{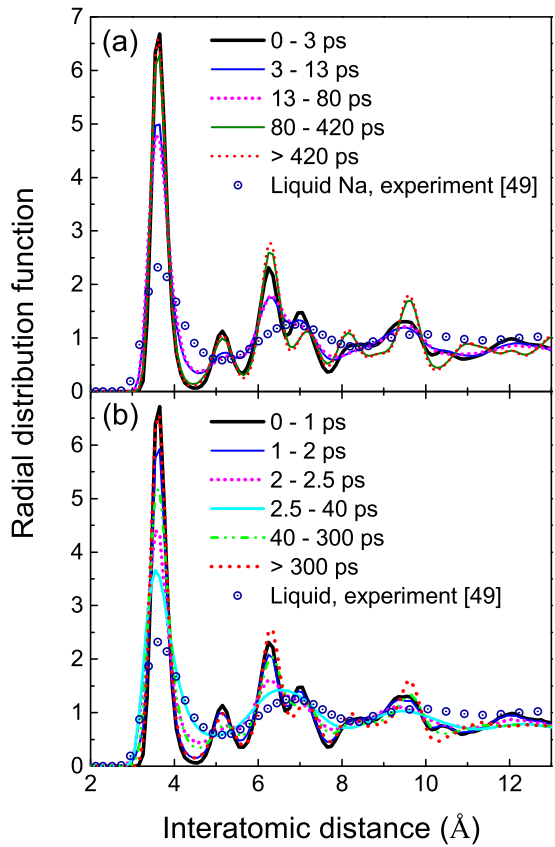}
\caption{
Radial distribution function (RDF) for the Na$_{1067}$ nanoparticle deposited at (a) 0.0068~eV/atom and (b) 0.068 eV/atom. 
Different curves describe the RDF evaluated over different time periods as indicated. These time periods are marked by dash-dotted vertical lines in Fig.~\ref{fig_Temp}(a,b).
}
\label{fig_gr}
\end{figure}

Further insights into the change of internal structure of the deposited nanoparticle can be drawn from the analysis of the radial distribution function (RDF). 
Figure~\ref{fig_gr} shows RDFs for the nanoparticle equilibrated at 77~K and deposited at $E_{\textrm{dep}} = 0.0068$~eV/atom (panel (a)) and $E_{\textrm{dep}} = 0.068$~eV/atom (panel (b)).
The plotted RDFs were averaged over different periods of time, which are marked by dashed vertical lines in Fig.~\ref{fig_Temp}(a,b).
The RDFs for the deposited Na$_{1067}$ nanoparticle are compared with the experimentally determined distribution for liquid sodium \cite{Murphy1973}.

\begin{sloppypar}
As shown in Fig.~\ref{fig_gr}(a) main peaks in the RDFs remain from the initial stage of deposition (within the first 3~ps of simulation) up to the last stage when the nanoparticle has reached thermal equilibrium with the substrate ($>420$~ps). Moreover, at all stages of the simulated trajectory the calculated RDFs differ from the one for liquid sodium (see open symbols). 
The melting phase transition and subsequent re-crystallization of the nanoparticle is clearly seen by the variation of RDFs in Fig.~\ref{fig_gr}(b).
Temperature of the nanoparticle deposited at 0.068~eV/atom reaches the maximum value at about 2.5~ps and it remains higher than the melting temperature up to about 40~ps, see Fig.~\ref{fig_Temp}(b). In between these time instances the nanoparticle transforms into a liquid droplet as confirmed by the close similarity of the calculated RDF with the one for liquid sodium \cite{Murphy1973}. At larger time instances (the region of $40 - 300$~ps) the nanoparticle re-crystallizes and eventually reaches thermal equilibrium with the substrate at about 400~ps. Re-crystallization manifests itself in reappearance of peaks in the RDF, which resembles the RDF in the very beginning of the simulation.
\end{sloppypar}

\section{Conclusion}
\label{sec:conclusion}

\begin{sloppypar}
In this paper deposition of a nanometer-sized sodium nanoparticle, containing 1067 atoms, on a MgO substrate was studied by means of molecular dynamics simulations using the MBN Explorer and MBN Studio software packages. We focused on the broad deposition energy range of $0.0034 - 1.36$~eV/atom, which covers both the soft-landing and nanoparticle fragmentation regimes. 
Simulations were performed 
at two different temperatures, 77~K and 300~K, at which the nanoparticle is either in the solid state or it forms a liquid droplet.
\end{sloppypar}

A force field describing the interaction of sodium atoms with the MgO surface was developed and used in the simulations. The force field was validated via calculating adsorption energies of a single Na atom on different MgO sites. The presented results agreed with the results obtained previously in \textit{ab initio} QM/MM simulations \cite{Bar2008,Bar2009,Dinh2010}. 

The process of nanoparticle deposition and subsequent relaxation on the surface has been studied in detail as a function of deposition energy. In particular, variation of the nanoparticle shape as a function of simulation time, its wetting properties, as well as the energy transfer between the nanoparticle and the substrate were analyzed on the timescale of up to several hundreds picoseconds. This study provides detailed insights into the dynamics of sodium nanoparticle deposition on MgO substrates which complement the information already gathered for small clusters from QM/MM simulations and which may be useful for experimental studies.
A similar analysis for other metallic aggregates deposited onto experimentally relevant oxide surfaces might reveal atomistic-level insights into the structure and shape of the deposited metal systems, which might be useful for technological applications.

\section*{Acknowledgements}

\begin{sloppypar}
This work was supported in part by Deutsche Forschungsgemeinschaft (Project no.~415716638);
by the European Union’s Horizon 2020 research and innovation programme -- the Radio-NP project (GA 794733) within the H2020-MSCA-IF-2017 call and the RADON project (GA 872494) within the H2020-MSCA-RISE-2019 call; by the Spanish Ministerio de Ciencia e Innovaci\'{o}n and the European Regional Development Fund (Project no. PGC2018-096788-B-I00); by the Fundaci\'{o}n S\'{e}neca -- Agencia de Ciencia y Tecnolog\'{i}a de la Regi\'{o}n de Murcia (Project No. 19907/GERM/15); and by the Conselleria d'Educaci\'{o}, Investigaci\'{o}, Cultura i Esport de la Generalitat Valenciana (Project no. AICO/2019/070).
PdV gratefully acknowledges the Alexander von Humboldt Foundation/Stiftung and 
the Spanish Ministerio de Ciencia e Innovaci\'{o}n for their financial support by means of, respectively, Humboldt (1197139) and Juan de la Cierva (FJCI-2017-32233) postdoctoral fellowships.
The possibility to perform computer simulations at Goethe-HLR cluster of the Frankfurt Center for Scientific Computing and the Scientific Computing Service of the University of Murcia is gratefully acknowledged.
\end{sloppypar}

\bibliography{library}

\end{document}